\documentclass[aps,prd,reprint,nofootinbib,onecolumn]{revtex4-2} 
\usepackage{color}
\usepackage{amsmath}
\usepackage{graphicx}
\usepackage{dcolumn}
\usepackage{bm}
\usepackage{amsmath}   
\usepackage{amssymb} 
\usepackage{graphics}
\usepackage{subfigure}
\usepackage[utf8]{inputenc} 

\begin{document}

\title{Pressure-Energy Equations of State of the Nucleon}

\vspace{0.6cm}


\author{Keh-Fei Liu }

\affiliation{
{\it
Department of Physics and Astronomy, University of Kentucky, Lexington, Kentucky 40506, USA} \\
{\it
Nuclear Science Division, Lawrence Berkeley National Laboratory, Berkeley, California 94720, USA} \\
}

\begin{abstract}
The pressure-energy equations of state in the nucleon are derived from the gravitational form factors, which parametrize matrix elements of the energy-momentum tensor (EMT), together with EMT conservation. The pressure and energy densities naturally separate into two distinct components. The static pressure distribution, arising from the Lorentz trace part of the EMT, as manifested in the spatial stress $\frac{1}{3} T^{ii}$, is equal to minus the corresponding trace part of the energy density. This relation may be interpreted as a consequence of the full or partial depletion of the gluon and quark condensates through the pressure-volume relation. The resulting  trace anomaly- and sigma term-induced pressure is shown, through its volume scaling, to play a pivotal role in QCD confinement. In contrast, the dynamic pressure distribution, associated with the traceless part of the stress tensor equals $1/d$ of the corresponding traceless part of the energy density, where $d$ is the spatial dimension. These static and dynamic components together satisfy the pressure-balance condition in the nucleon.  We further show that the same pressure-energy equations of state hold for vortices in type-II superconductors, where the static pressure-energy relation originates from the depletion of the Cooper-pair condensate. Moreover, these equations of state are identical to those of the $\Lambda{\rm CDM}$ cosmological model, where the static pressure-energy relation is generated by the cosmological constant.

\end{abstract}

\maketitle

\section{Introduction} \label{intro}

 A comprehensive understanding of how QCD gives rise to linear confinement in
heavy quarkonium--its origin and underlying dynamics--remains a central and longstanding theoretical challenge. It is intuitively related to the notions of pressure and force. The energy-momentum tensor (EMT) is the fundamental operator for characterizing mass, energy, and pressure in quantum field theory. In particular, the gravitational form factors (GFFs) of the quark and gluon components of the EMT in hadrons can be accessed from generalizes parton distribution (GPD) analysis with model-dependent extractions and are calculable in lattice QCD.

Considerable recent interest has focused on the pressure distributions inside hadrons.  The 
$D$-term, one of the gravitational form factors of the energy-momentum tensor, is accessible through  GPDs~\cite{Ji:1996nm,Diehl:2003ny} measured in hard exclusive processes. The Fourier transform of the 
$D$-term can be interpreted as the internal pressure distribution of hadrons, in close analogy with that of a fluid
~\cite{Polyakov:2002yz,Polyakov:2018zvc}. 
The energy and pressure distributions of quarks and gluons derived from the energy-momentum tensor are analyzed within the instant and front forms of relativistic dynamics. The separate quark and gluon contributions are obtained in the Breit, elastic, infinite-momentum, and Drell-Yan frames~\cite{Lorce:2018egm,Lorce:2025oot}.
Experimental access to the quark and gluon 
$D$-form factors is provided through GPDs, which can be extracted from deeply virtual Compton scattering (DVCS)
~\cite{Burkert:2018bqq,Burkert:2021ith,Duran:2022xag,Goharipour:2025lep}. Extensive theoretical works followed. These include formal discussions of mass, energy, and pressure decompositions~\cite{Lorce:2021xku,Liu:2021gco,Liu:2023cse}, transverse momentum~\cite{Lorce:2023zzg}, momentum currents~\cite{Ji:2021mfb,Ji:2025gsq}, twist-4 $\bar{C}_{\rm q,g}$ form factors~\cite{Tanaka:2022wzy}, and large-momentum behavior~\cite{Tong:2021ctu}. Heavy quarkonium productions near threshold~\cite{Hatta:2019lxo,Hatta:2021can,Guo:2023pqw,Guo:2023qgu,Hatta:2025ryj} have been considered a viable way to extract the glue trace anomaly form factor. The gravitational form factors have been approached with chiral theories~\cite{Freese:2019eww,Alharazin:2020yjv,Owa:2021hnj,Won:2023zmf,Stegeman:2025tdl,Stegeman:2025sca}, holographic formalism~\cite{Mamo:2019mka,Mamo:2021krl,Mamo:2022eui,Fujita:2022jus}, light-front formalism~\cite{Chakrabarti:2020kdc,Freese:2021czn,Panteleeva:2021iip,Kim:2021jjf,Choudhary:2022den},
Schwinger-Dyson equations~\cite{Yao:2024ixu}, the quark-soliton model~\cite{Won:2022cyy}, the MIT bag model~\cite{Neubelt:2019sou},
and skyrmion~\cite{Fujii:2025aip}. In particular, the behaviors of the quark and gluon 
$D$-form factors in lattice calculation are consistent with those extracted from experiments, although the extraction is model dependent and statistically limited and  the lattice results are not yet at the continuum limit~\cite{Shanahan:2018nnv,Hackett:2023rif}.

\section{Mass and Rest Energy} \label{mass}

The mass, energy, and pressure of hadrons are encoded in the EMT, constructed from quark and gluon operators in the symmetric, gauge-invariant Belinfante form,
\begin{equation}
    T^{\mu\nu} = \sum_f \frac{i}{4}\bar{\psi}_f \gamma^{\{\mu}\!\stackrel{\leftrightarrow}{D}\!{}^{\nu\}}\psi_f - G^{\mu\lambda}G^{\nu}_{\lambda} + \frac{1}{4} g^{\mu\nu}G^{\alpha\beta}G_{\alpha\beta},
\end{equation}
where the color indices are summed and not shown explicitly. The quark sector contains a sum over quark flavors $f$. The hadron mass is determined by the trace of the EMT, which in QCD is known to exhibit a quantum (trace) anomaly after renormalization
~\cite{Chanowitz:1972vd,Crewther:1972kn,Chanowitz:1972da,Collins:1976yq}, 
\begin{equation} \label{trace}
T^{\mu}_{\mu} = T_{a\,\, \mu}^{\mu} + T_{q\,\, \mu}^{\mu},
\end{equation}
where $T_{a\,\, \mu}^{\mu}$ denotes the trace anomaly contribution and $T_{q\,\, \mu}^{\mu}$ the quark sigma term contribution. They have the expressions
\begin{eqnarray}  \label{eq:trace_q,g}
T_{a\,\, \mu}^{\mu} = \frac{\beta(g)}{2g} G^{\alpha\beta} G_{\alpha\beta} 
\label{trace_g} + \sum_f \gamma_m (g)\, m_f \bar{\psi}_f \psi_f, \hspace{1cm}
T_{q\,\, \mu}^{\mu} = \sum_f m_f \bar{\psi}_f \psi_f,  
\end{eqnarray}
and both are renormalization group invariant. Thus,
\begin{equation} \label{invariant_M}
M=\frac{\langle P| \int d^3  \vec{x}\, \gamma T^{\mu}_{\mu}(x)|P\rangle}{\langle P|P\rangle}
= \frac{\langle P| \int d^3  \vec{x}\, \gamma (T_{a\,\, \mu}^{\mu}(x) + T_{q\,\, \mu}^{\mu}(x))|P\rangle}{\langle P|P\rangle}
\end{equation}
where $\gamma = 1/\sqrt{1 - (v/c)^2}$ is the factor in Lorentz transformation. This demonstrates that the hadron mass is scale and frame independent, as expected for a Lorentz scalar. Lattice QCD calculations~\cite{Yang:2018nqn} show that the trace anomaly provides the dominant contribution to the nucleon mass ($\sim$ 91\%), while the strange and $u + d$ sigma terms from disconnected insertions contribute $\sim$ 6\%. The remaining $\sim$ 3\% arises from the $u+d$ valence contribution in the connected 
insertion~\cite{Liu:2021gco,Liu:2023cse}. 

On the other hand, the decomposition of the rest energy follows from the Hamiltonian \\
\mbox{$ H = \int d^3\vec{x}\,\, T^{00}(x)$.} As a symmetric rank-two tensor, the QCD energy-momentum tensor can be decomposed into a traceless part associated with scale-symmetric dynamics and a trace part associated with explicit and anomalous breaking of scale symmetry~\cite{Ji:1994av,Ji:1995sv}
\begin{equation} \label{trace_separation}
T^{\mu\nu} = \overline{T}^{\mu\nu} + \frac{1}{d+1} \, g^{\mu\nu} T_{\rho}^{\rho}.
\end{equation}
where $d$ is spatial dimensions. Up to this point, the separation is scale and scheme independent. However, the traceless part 
$\overline{T}^{\mu\nu}$ can be further decomposed into quark and gluon contributions, which are individually scale dependent. In this case, the rest energy $E_0$ can be written as the sum of the traceless and trace components~\cite{Ji:1994av,Ji:1995sv}.
\begin{equation}  \label{E-traceless}
E_0 = \overline{E}_0 + E_{0, {\rm tr}}
\end{equation}
where, 
 \begin{eqnarray}    \label{equ:E_0}
 \overline{E}_0 \!\! &=& \!\!\langle P|\Big[\int d^3\vec{x}\, (\frac{i}{4} \sum_f \bar{\psi}_f \gamma^{\{0}\!\stackrel{\leftrightarrow}{D}\!{}^{0\}}\psi_f 
 - \frac{1}{4} T_{q\, \mu}^{\mu}) + \int d^3\vec{x}\, \frac{1}{2} (B^2 + E^2)\Big]|P \rangle/\langle P|P \rangle \label{eq:traceless_E} \\
   &=& \!\! \frac{3}{4}\, [\langle x\rangle_q (\mu)+ \langle x\rangle_g (\mu)]\, M_N \nonumber \\
   E_{0, {\rm tr}}\!\! &=& \!\! \frac{1}{4}\, \langle P| \Big[\int d^3\vec{x}\,  T_{a\,\, \mu}^{\mu} 
+  \int d^3\vec{x}\,  T_{q\,\, \mu}^{\mu} \Big] |P\rangle/\langle P|P \rangle = \frac{1}{4} M_N  \label{eq:trace_E}
\end{eqnarray}  
As shown in the second line of Eq.~(\ref{eq:traceless_E}), the nucleon matrix elements of the traceless EMT $\overline{T}^{00}$ at rest, which encode the quark kinetic and potential energies and the gluon field energy, are proportional to $3/4$ of the second moments of the parton distribution functions obtained from the operator product expansion~\cite{Ji:1994av,Ji:1995sv,Yang:2018nqn}. These correspond to the quark and gluon momentum fractions on the light front, $\langle x\rangle_q(\mu)$ and $\langle x\rangle_g(\mu)$. They are scale dependent and require specification of a renormalization scheme, such as the $\overline{\rm MS}$ scheme.

In the rest frame, 3/4 of the rest energy $E_0$ arises from the traceless component $\overline{T}^{00}$, while the remaining 1/4 comes from the trace part~\cite{Ji:1994av,Ji:1995sv}. More generally, in d spatial dimensions, the trace part contributes 1/(d+1) of the rest energy, and the traceless part contributes the remaining d/(d+1). This partition is determined solely by the spatial dimension d. Since energy, unlike mass, is frame dependent, these results apply only in the rest frame.

\section{Gravitational Form Factors}   \label{sec:GFF}

The nucleon mass and rest energy can also be obtained from the forward limit of the GFFs of the EMT. The off-forward GFFs are encoded in GPDs~\cite{Ji:1996nm,Diehl:2003ny}, which are accessed experimentally through DVCS and deeply virtual meson production at JLab~\cite{Dudek:2012vr}, HERMES~\cite{HERMES:2008abz}, and COMPASS~\cite{COMPASS:2018pup}, and are calculable in lattice QCD~\cite{Liu:2021lke,Alexandrou:2020sml,Wang:2021vqy,Hackett:2023rif}. These form factors may provide insight on  the spatial distributions of mass, energy, and pressure in the nucleon.
They contain the following terms~\cite{Kobzarev:1962wt,Pagels:1966zza,Ji:1996ek}
for the quarks and gluons 
\begin{eqnarray} \label{GFF}
\langle P'| T_{q, g}^{\mu\nu}(0)|P\rangle &=& \bar{u}(P')\big[\frac{\gamma^{\{\mu}\bar{P}^{\nu\}}}{2} A_{q,g}(q^2) +
 \frac{i \bar{P}^{(\mu} \sigma^{\nu)\alpha} q_{\alpha}}{4M} B_{q,g}(q^2)  \nonumber \\ 
     &+&  \frac{q^{\mu}q^{\nu} - g^{\mu\nu}q^2}{M}D_{q,g}(q^2)+  M g^{\mu\nu} \overline{C}_{q,g}(q^2)\big] u(P).
\end{eqnarray}

In the forward limit, $A_{q,g}(0)= \langle x\rangle_{q,g}(\mu) $ gives the quark and gluon momentum fractions, while \\
\mbox{$J_{q,g}(0) = (A_{q,g} (0) + B_{q,g} (0))/2$} gives the corresponding angular momentum fractions~\cite{Ji:1996ek}. They obey the sum rules $A_q(0) + A_g(0) = 1$ and \mbox{$J_q(0) + J_g(0)  = 1/2$ }. 
%
By analogy with the stress tensor of a continuous medium, it has been suggested~\cite{Polyakov:2002yz,Polyakov:2018zvc} that the $D_{q,g}(q^2)$ form factors encode the shear forces and pressure distributions of quarks and gluons inside the nucleon. Similarly, $\overline{C}_{q,g}(0)$ has been identified with the pressure-volume work~\cite{Lorce:2017xzd,Lorce:2018egm,Liu:2021gco,Liu:2023cse}. This interpretation is consistent with the fact that $\overline{C}(0)$ equals the normal stress $T^{ii}(0)$~\cite{Liu:2021gco}.

From the GFFs, it was shown~\cite{Lorce:2017xzd} that the nucleon mass obtained from the trace reads $M = (\langle x\rangle_q (\mu)+ \langle x\rangle_g (\mu)) M + 4 (\bar{C}_q(0) + \bar{C}_g (0)) M$ whereas the rest energy is
$E_0 = (\langle x\rangle_q (\mu)+ \langle x\rangle_g (\mu)) M +  (\bar{C}_q(0) + \bar{C}_g (0)) M$. They apparently differ from the four-term decomposition in Eqs.~(\ref{eq:traceless_E}) and (\ref{eq:trace_E}) and have led to some controversy~\cite{Lorce:2017xzd,Metz:2020vxd,Lorce:2021xku,Ji:2021mtz}. Once  $\bar{C}_{q,g}(0)$ is identified through the forward matrix elements of the spatial stress $T^{ii}$, namely 
\begin{equation} \label{eq:C}
 (\bar{C}_q(0) + \bar{C}_g (0))M =
 -\frac{1}{3} \frac{\langle P| T^{ii}(0)|P \rangle}{\langle P|P\rangle} =
\frac{1}{4} (f_a +  f_{\sigma})M - \frac{1}{4} (\langle x\rangle_q + \langle x\rangle_g)M,
\end{equation}
where $f_a/f_{\sigma}$ denotes the anomaly and sigma-term fractions of the nucleon mass, respectively, one recovers Eqs.~(\ref{invariant_M}), (\ref{eq:traceless_E}), and (\ref{eq:trace_E})~\cite{Liu:2023cse}. 
Since $\bar{C}_q(q^2) + \bar{C}_g (q^2) = 0$ for all $q^2$ as a consequence of EMT conservation, $\partial_{\nu} T^{\mu \nu}=  0$,
these terms are often discarded. Doing so yields the inconsistent decomposition $M = E_0 = (A_q(0) + A_g(0))M$, instead of Eqs.~(\ref{invariant_M}), (\ref{eq:traceless_E}), and (\ref{eq:trace_E}). Therefore, the expression for $\bar{C}_q(0) + \bar{C}_g (0)$ 
should be retained, rather than set to zero, to maintain consistency between the Lorentz- and CPT-covariant GFF decomposition and the decomposition based on the irreducible components of the EMT.

Beyond restoring consistency in the mass and energy decompositions,  $\overline{C}(0) = \overline{C}_q(0) + \overline{C}_g(0)$, being related to the spatial diagonal component $T^{ii}$ of the EMT, admits an interpretation as the total pressure-volume work~\cite{Lorce:2017xzd,Lorce:2021xku,Liu:2021gco,Liu:2023cse}. The fact that $\overline{C}(0) = 0$ implies that the net total pressure vanishes, reflecting a balance between the quark kinetic pressure and gluon radiation pressure on the one hand and the trace contribution of the EMT on the other. This relation is analogous to the virial theorem for a nonrelativistic bound state~\cite{Liu:2023cse}, in which the virial of the kinetic energy is balanced by that of the potential, i.e., the spatial moment of the force (i.e., $\vec{r}\cdot \vec{F}$)~\cite{Maranganti:2010,Rossi:2010}. We should note that $\overline{C}_q$ and $\overline{C}_g$ do not vanish separately. 
To gain deeper insight into the role of pressure, one must examine its local distribution.

\section{Definition of pressure}

It is important to distinguish between two notions of pressure. The first is the thermodynamic pressure, defined for a system in thermal equilibrium that is homogeneous and extensive, $P = k_B T (\frac{\partial \ln Z}{\partial V})_{N,T}$, where $Z$ is the partition function. The second is the mechanical pressure, defined through the spatial components of the energy-momentum tensor, $P = \frac{1}{3} \langle T^{ii}\rangle$,
with the EMT given by $\langle T^{\mu\nu} (x)\rangle = \frac{2}{\sqrt{-g}} \frac{\delta \ln Z}{\delta g_{\mu\nu}(x)}$. In systems such as an ideal gas, perfect fluid, or plasma—where thermal equilibrium, homogeneity, and isotropy hold—the two definitions coincide. The same is true in finite-temperature QCD once the trace anomaly is properly incorporated into the EMT~\cite{Drummond:1999he}.

Although a neutron star is globally inhomogeneous, it can be described by local thermal equilibrium because the microscopic mean free path is much shorter than the scale of density variation, validating the local density approximation. 
The situation is markedly different for the nucleon. As an eigenstate of the Hamiltonian, it carries no entropy and does not admit a thermodynamic description. Moreover, the interaction range of quarks and gluons is comparable to the nucleon size, precluding local thermodynamic equilibrium. Consequently, a thermodynamic pressure cannot be defined. 
Nevertheless, a local mechanical pressure can be defined through the nucleon matrix elements of the energy-momentum tensor with curvilinear coordinates. The EMT is obtained from the variation of the action with respect to the metric as described in the last paragraph. In the nucleon, the variation of the metric is with respect to the nucleon two-point function
\begin{equation}
    \frac{\langle N|T^{\mu\nu}(r)|N\rangle}{\langle N|N\rangle} = \lim_{\tau 
 \rightarrow \infty} \frac{2}{\sqrt{-g}\,\tau}\frac{\delta \ln C_2(\tau,g)}{\delta g_{\mu\nu}(r)},
\end{equation}
where $C_2(\tau,g)_ {\stackrel{\longrightarrow}{\tau \rightarrow \infty}}  |Z|^2 e^{-E_N \tau}$ denotes the nucleon two-point correlation function in the Euclidean path-integral formalism under the action $S_{\rm QCD} = \int d^4x \,\sqrt{-g}\,\,\mathcal{L}_{\rm QCD}(x)$.
%

Noting that an isotropic spatial variation of the metric 
$g_{ij}(x)$ corresponds to a local variation of the spatial volume, $\delta v(r) = \delta (\sqrt{-g(r)})$~\cite{Weinberg:1972kfs}, the mechanical pressure is 
%
\begin{equation}   \label{eq:p_mech}
    p_{\rm mech} (r) = \frac{1}{3} \frac{\langle N|T^{ii}(r)|N\rangle}{\langle N|N\rangle} = - \frac{\delta E_N}{\delta v(r)},
\end{equation}
The last term is the normal stress in a continuous medium~\cite{Landau:1986aog}. As we shall show, it plays a central role in elucidating the origin of the trace-anomaly contributions to the energy density and pressure distributions in the nucleon, as well as the volume scaling of the different components of energy and pressure.
In the following, we focus exclusively on the mechanical pressure 
$p_{\rm mech}$, which we denote simply by $p$.

\section{Off-diagonal EMT matrix elements} 
\label{off}

It has been suggested that the mechanical pressure can be extracted from the gravitational form factors of the energy-momentum tensor $T^{ii}$~\cite{Polyakov:2002yz,Polyakov:2018zvc}.
In the Breit frame, the normalized off-diagonal EMT matrix elements $\langle\langle  T^{\mu\nu}(0)\rangle\rangle = \langle P',s|T^{\mu\nu}|P,s\rangle/2E$ are 
\begin{eqnarray}
  \langle\langle T^{\mu}_{\mu}\rangle\rangle (q^2) &=& M\big[A(q^2) + \frac{q^2}{4M^2}B(q^2) -3 \frac{q^2}{M^2} D(q^2) + 4\overline{C}(q^2)\big] \label{eq:traceFF}\\
 \langle\langle T^{00}\rangle\rangle (q^2) &=& M \,\big[ A(q^2) + \frac{q^2}{4M^2}B(q^2) - \frac{q^2}{M^2}D(q^2) + \overline{C}(q^2)\big] \label{eq:EnergyFF}\\
 \langle\langle T^{ij}\rangle\rangle (q^2) &=& M\Big[ (q^iq^j + \frac{1}{3}\delta_{ij}q^2) \frac{D(q^2)}{M^2} + \frac{2}{3}\delta_{ij} \frac{q^2D(q^2)}{M^2} - \delta_{ij} \overline{C}(q^2) \Big]. \label{eq:TijFF}
 \end{eqnarray}
where $\langle\langle T^{ij}\rangle\rangle$ is decomposed into traceless and trace components with respect to the spatial indices. In addition, as shown in Eqs.~(\ref{trace}) and (\ref{eq:trace_q,g}), the off-forward matrix elements of the trace, $\langle\langle T^{\mu}_{\mu}\rangle\rangle$, can be expressed in terms of the combined trace-anomaly and quark sigma-term operators. We, therefore, define the mass form factor from the total trace operator as
\begin{equation} \label{eq:G_m}
\langle\langle T^{\mu}_{\mu}\rangle\rangle (q^2) = M \,G_m (q^2) 
\end{equation}
From now on, we shall consider the combined quark and glue form factors, so that there is no scale dependence.

It has been noted~\cite{Ji:2021mtz,Liu:2023cse} that, since $\overline{C}(q^2)=0$ as a consequence of EMT conservation, Eq.~(\ref{eq:traceFF}) is equal to Eq.~(\ref{eq:G_m}). Choosing $A(q^2), B(q^2)$ and $G_m(q^2)$ as independent form factors, whose forward limits satisfy $A(0)= G_m(0) = 1$ and $B(0)=0$, the $D(q^2)$ form factor can be expressed~\cite{Liu:2023cse,Mamo:2021krl} as~\footnote{We thank F. Yuan for pointing out that $D_q$ and $D_g$
 can be determined separately in experiments and lattice QCD. In that case, one may introduce $f_{q,g}(q^2,\mu)$ such that
\mbox{$D_{q,g}(q^2,\mu)  = \frac{1}{3} \Big [\frac{(A_{q,g}(q^2,\,\mu) - f_{q,g} (q^2,\,\mu)\,G_m(q^2)+ 4\overline{C}_{q,g}(q^2,\,\mu))M^2 }{q^2} + \frac{B_{q,g}(q^2, \mu)}{4} \Big ]$,} where $f_q(q^2,\mu) + f_g(q^2,\mu) = 1$.  } 

\begin{equation}   \label{eq:D(q^2)}
D (q^2) = \frac{1}{3} \Big [\frac{(A(q^2) - G_m(q^2))M^2 }{q^2} + \frac{B(q^2)}{4} \Big ].
\end{equation}
In this case, $D(0)$, the so-called D-term, is expressed in terms of the square radii of the $A$ and $G_m$ form factors,
\begin{equation}  \label{eq: D0}
D(0) = \frac{1}{18} \big[\langle r^2\rangle_{A} -\langle r^2\rangle_m \big]M^2 
\end{equation}
where $\langle r^2\rangle = - \frac{6}{F(0)}  \frac{d F(q^2)}{d q^2}|_{q^2=0}$ for the form factor $F(q^2)$. 
As such, the D-term does not need to be regarded as a fundamental parameter, as sometimes suggested~\cite{Polyakov:2018zvc}. Rather, it can be expressed as the difference between the mean-square radius $\langle r^2\rangle_{A}$
associated with $A(q^2)$ and the mean-square radius $\langle r^2\rangle_m$ of the mass form factor $G_m(q^2)$. Equivalently, it may be written as the difference between the radii extracted from the trace form factor $\langle\langle T^{\mu}_{\mu}\rangle\rangle (q^2)$ in Eq.~(\ref{eq:G_m}) and the energy form factor $\langle\langle T^{00}\rangle\rangle (q^2)$  in Eq.~(\ref{eq:EnergyFF})~\cite{Ji:2021mtz}\!\cite{Fujii:2025aip}. These radii are experimentally accessible and calculable in lattice QCD. Recent lattice studies~\cite{Shanahan:2018nnv,Hackett:2023rif} find $D(0)$ to be negative, implying that the mass distribution from the trace anomaly and sigma terms is more spatially extended than the quark and gluon energy distributions reflected in $\langle r^2 \rangle_A$.

Denoting $\langle\langle T^{00}\rangle\rangle (q^2)$ as the energy form factor $E (q^2)$, it can be decomposed into trace and traceless components corresponding to the operator $T^{00}$. Substituting $D(q^2)$ from Eq.~(\ref{eq:D(q^2)}), one obtains

\begin{equation}
    E(q^2) \equiv \langle\langle T^{00}\rangle\rangle (q^2) = \overline{E}(q^2) + E_{\rm tr}(q^2),
\end{equation}
where,
\begin{eqnarray}
    \overline{E}(q^2) &=& M\,\big[\frac{2}{3}A(q^2)+ \frac{q^2}{6M^2}B(q^2) +\frac{1}{12}G_m(q^2)\big],  \label{eq:Ebar} \\
E_{\rm tr}(q^2) &=& \frac{M}{4} G_m(q^2). 
\label{eq:Etr}
\end{eqnarray}

On the other hand, the total pressure form factor, $P(q^2) = \frac{1}{3} \langle\langle T^{ii}\rangle\rangle = \frac{2}{3} M q^2 D(q^2)$, can be decomposed analogously into traceless and trace components 
%
\begin{eqnarray}
    \overline{P}(q^2) &=& M\big[\frac{2}{9}A(q^2) + \frac{q^2}{18M^2} B(q^2) + \frac{1}{36} G_m(q^2)],  \label{eq:pbar}\\
   P_{\rm tr} (q^2) &=& -\, \frac{M}{4} G_m(q^2). \label{eq:ptr}
\end{eqnarray}
In the forward limit, these reduce to the corresponding terms in the forward matrix element of $T^{ii}$
in Eq.~(\ref{eq:C}). From Eqs.~(\ref{eq:Ebar}, \ref{eq:Etr}, \ref{eq:pbar}, \ref{eq:ptr}), one observes that the pressure and energy form factors are related as follows:

\vspace{-0.5cm}
\begin{eqnarray}
   \overline{P}\,(q^2) &=& \frac{1}{d} \, \overline{E}\,(q^2), \label{eq:PEq2_bar} \\
P_{\rm tr}(q^2) &=& -\, E_{\rm tr}\,(q^2).
\label{eq:PEq2_tr}
\end{eqnarray}
In the forward limit, they reproduce the corresponding relations between the $T^{ii}$ and $T^{00}$ matrix elements for both the traceless and trace parts in Eqs. (\ref{eq:traceless_E}), (\ref{eq:trace_E}) and (\ref{eq:C}).

\subsection{Spatial distribution}

It is customary in the literature to assume that, in the Breit frame where the energy transfer vanishes, the spatial distribution is obtained from the three-dimensional Fourier transform of the relevant form factor, as in the nonrelativistic case,

\begin{equation}
    \widetilde{F}(r) = \int \frac{d^3 q}{(2\pi)^3}\, e^{-i \vec{q}\cdot \vec{r}} \,F(q^2)|_{q^0 = 0}.
\end{equation}
However, it has been pointed out~\cite{Burkardt:2000za, Miller:2007uy, Miller:2010nz, Miller:2018ybm, Jaffe:2020ebz} that this definition of a time-independent three-dimensional density is not valid for systems with relativistic constituents. The main issue is that the coordinate appearing in the Fourier transform is not defined relative to the position of the hadron itself. To remedy this, one introduces a localized wave packet to describe the hadron state. In this framework, the hadron size $\Delta$, the spatial width of the wave packet $R$, and the Compton wavelength $1/M$ must satisfy the hierarchy~\cite{Jaffe:2020ebz}
\begin{equation}
1/M \ll R \ll \Delta  
\end{equation}
so that relativistic localization effects are suppressed while the wave packet remains small compared to the intrinsic size of the hadron.  
This condition is well satisfied for atoms and nuclei, where $\Delta$ exceeds $1/M$ by many orders of magnitude. In contrast, for the nucleon, $\Delta \sim 0.8$ fm,
while $1/M_N \sim 0.2$ fm. There is, therefore, insufficient separation of scales to choose a wave packet size $R$ satisfying this double inequality, and the resulting density inevitably retains sensitivity to the choice of $R$~\cite{Jaffe:2020ebz}.

In lattice QCD, the spatial density can be computed directly. The absence of an intrinsic reference point for the hadron position can be resolved by inserting conserved point-split charge operators or point operators with a normalization factor  $Z_V$
 on quark lines other than the one carrying the local operator and on the same time slice. By the Ward identity, these insertions leave the quark propagators unchanged while introducing spatial coordinates $x_i$ associated with the spectator quarks. These coordinates define an unambiguous reference point for the nucleon position. The resulting construction yields a spatial density independent of the Breit frame Fourier transform and provides a direct test of the approximation based on form factors. Details will be reported elsewhere~\cite{Liu:2026}.

Although the three-dimensional Fourier transform of nucleon form factors does not provide a rigorously defined spatial density, it may still serve as a reasonable approximation for the proton in cases where the distribution has the same sign and the mean-square radius, obtained from the slope of the form factor at $q^2 = 0$, characterizes the system’s size. (The neutron electric form factor is a well-known exception.) With this caveat, we proceed to extract spatial distributions from the Fourier transform in the Breit frame in the present work. 
 
In terms of spatial distributions, the energy density can be decomposed as $\epsilon (r) = \overline{\epsilon} (r) + \epsilon_{\rm tr} (r)$, where $\overline{\epsilon} (r)$ and $\epsilon_{\rm tr} (r)$ denote the traceless and trace contributions, respectively. These are expressed in terms of the coordinate-space distributions $\widetilde{A}(r), \widetilde{B}(r)$ and $\widetilde{G}_m(r)$
\begin{eqnarray}
\overline{\epsilon}(r)&=& M\,(\frac{2}{3}\widetilde{A}(r)+ \frac{1}{3} \widetilde{B}(r) + \frac{1}{12}\widetilde{G}_m(r)), \\
\epsilon_{\rm tr}(r)&=& \frac{M}{4} \widetilde{G}_m(r).
\end{eqnarray}
Similarly, the pressure is decomposed as $p(r) = \overline{p} (r) + p_{\rm tr} (r)$, where the traceless and trace components are given by
\begin{eqnarray}
 \overline{p}(r)&=& M\,(\frac{2}{9}\widetilde{A}(r)+ \frac{1}{18} \widetilde{B}(r) +\frac{1}{36}\widetilde{G}_m(r)), \label{eq:pr_overline}\\
p_{\rm tr}(r) &=& - \frac{M}{4} \widetilde{G}_m(r). \label{eq:prtr}
\end{eqnarray}
From these expressions, the pressure-energy equations of state follow directly:
\begin{eqnarray}
    p_{\rm tr}(r)&=& - \,\epsilon_{\rm tr}(r) \label{eq:petr} \\
    \overline{p}(r) &=& \frac{1}{d}\, \overline{\epsilon}(r) \label{eq:pe-overline}
\end{eqnarray}
The direct lattice evaluation of the spatial distributions from the corresponding  $T^{00}$ and $T^{ii}$ operators, using the prescription described above, yields the same equations of state. 

 These equations of state have been obtained previously by examining the twist-2 and twist-4 gravitational form factors~\cite{Won:2022cyy}.  
 In this work, we investigate the underlying physical origins of them, analyze their volume-scaling properties, and compare them with analogous equations of state in other physical systems.

\begin{figure}[htbp]     \centering
{\includegraphics[width=0.4\hsize]{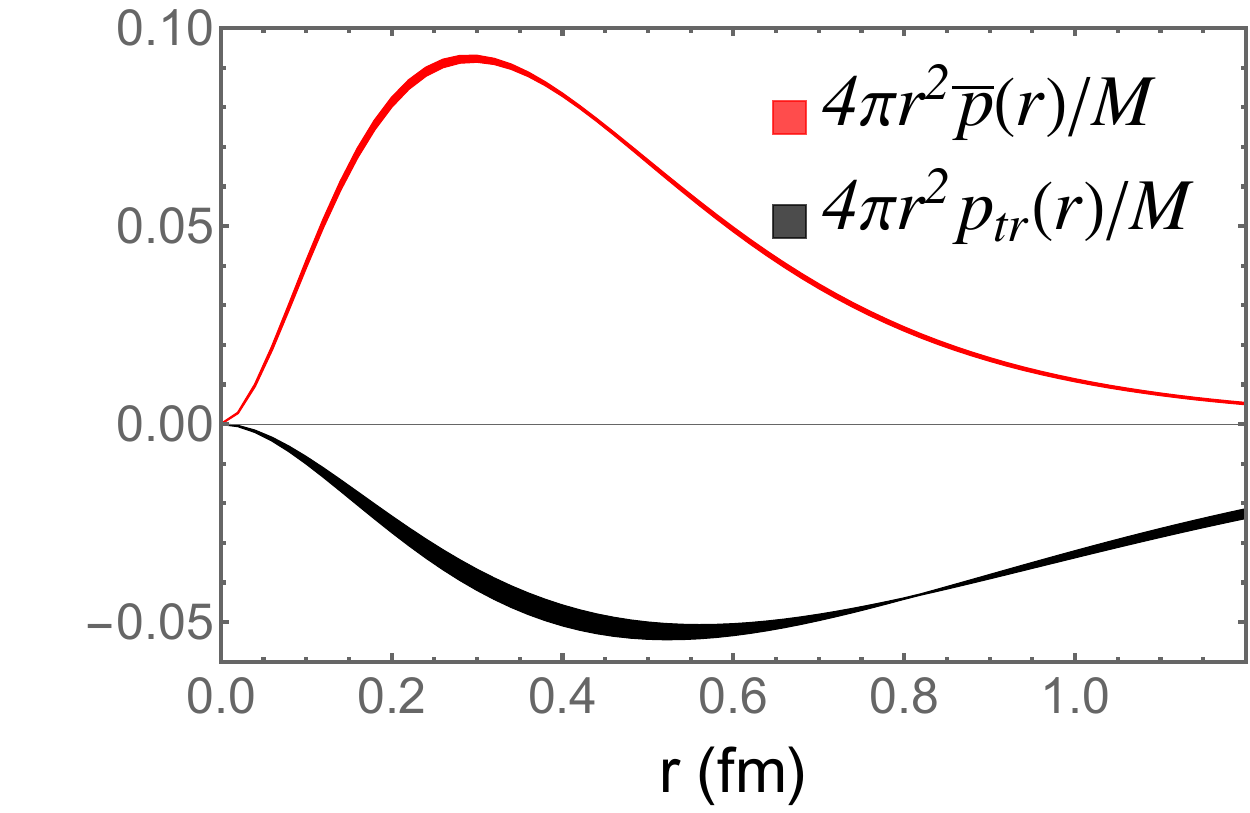} }
{\includegraphics[width=0.4\hsize]{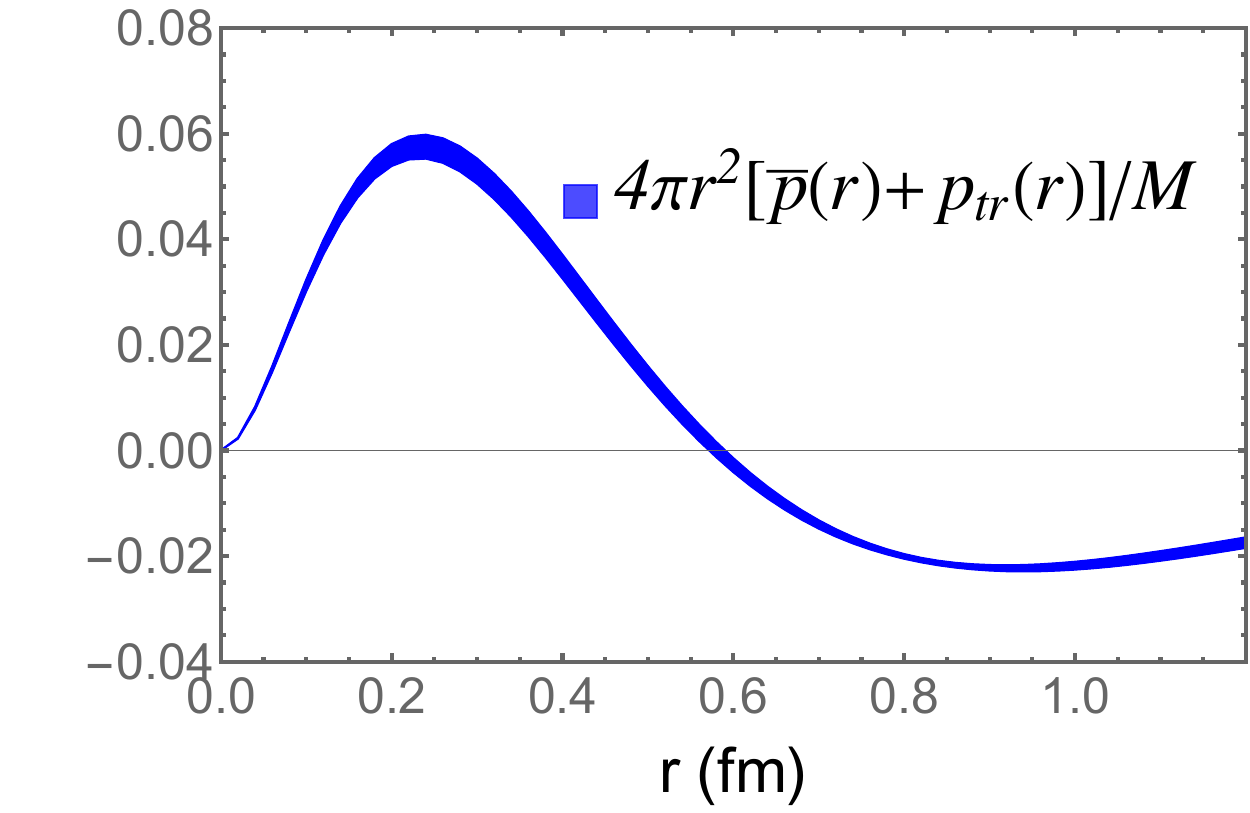} } 
\caption{
Left panel: $4\pi r^2\, \overline{p}(r)/M$ and $4\pi r^2\, p_{\rm tr}(r)/M$ plotted separately as functions of $r$. Right panel: the total $4\pi r^2\, p(r)/M$ as a function of $r$. The latter agrees with the result obtained from the lattice calculation of $\widetilde{D}(r)$~\cite{Hackett:2023rif}.  \label{fig:pr}
}
 \end{figure}

Before proceeding, we examine the total pressure distribution. It was found in the lattice calculation~\cite{Shanahan:2018nnv} that the total pressure distribution derived from
$p(r)/M =  \frac{2}{3}\frac{1}{M^2}\frac{1}{r^2}\frac{d}{d r}\!\left(r^2\frac{d\mathcal\, \widetilde{D}(r)}{d r}\right)$ exhibits a sign change, with negative pressure at larger distances, thereby ensuring the mechanical stability of the nucleon. 
This sign change can be understood from the combined behavior of $\overline{p}(r)$ and $p_{\rm tr}(r)$. Using lattice results~\cite{Hackett:2023rif} for $\widetilde{A}(r)$ and $\widetilde{G}_m(r)$ in Eqs.~(\ref{eq:pr_overline}) and (\ref{eq:prtr}), we plot $\overline{p}(r)/M$ and
$p_{\rm tr}(r)/M$ in Fig.~\ref{fig:pr}. The contribution from $\widetilde{B}(r)$ is neglected since $B(0) = 0$, and a quenched lattice calculation shows that both $B_{q}(q^2)$ and $B_g(q^2)$ are nearly flat in the region $|q^2| < 2\, {\rm GeV^2}$~\cite{Deka:2013zha}.

The combined pressure $p(r) =\overline{p}(r)+ p_{\rm tr}(r)$, shown in the right panel of Fig.~\ref{fig:pr}, coincides with that obtained from $\widetilde{D}(r)$ in lattice calculations~\cite{Shanahan:2018nnv, Hackett:2023rif}. It has been noted that from the von Laue stability condition~\cite{Laue:1911lrk}
\begin{equation} \label{eq:vonLaue}
   \int d^3 r\, p(r) = 0, 
\end{equation}
there should be a node in $p(r)$. Such a node structure has been predicted in various models, including the chiral quark model~\cite{Goeke:2007fp} and the Skyrmion model~\cite{Fujii:2025aip}, and has also been inferred from a combination of lattice QCD calculations and experimental data~\cite{Ji:2025qax}.

The von Laue stability condition admits a clear physical interpretation. The spatial integral of 
$p(r)$ corresponds to the forward matrix element: 
$\int d^3 r\, \overline{p}(r) = \frac{1}{4}\, A(0)$ and $\int d^3 r\, p_{\rm tr}(r) =  - \frac{1}{4}\, G_m(0)$~\cite{Liu:2023cse}, as seen from Eq.~(\ref{eq:C}). With $A(0) = G_m(0) =1$, the condition reflects a balance between the positive traceless contribution and the negative trace component.
Since $\langle r^2\rangle_{A} < \langle r^2\rangle_m$, this implies $D(0) < 0$ from 
 Eq.~(\ref{eq: D0}), the inner region is dominated by the repulsive traceless pressure, while the outer region is governed by the attractive trace part, ensuring mechanical stability.

The traceless pressure $\overline{p}_{\rm tr}$ originates from the quark kinetic and potential energies and the gluon radiation pressure encoded in the $\overline{T}^{ii}$ matrix elements. It is positive and satisfies $\overline{p}_{\rm tr} = \frac{1}{3}\overline{\epsilon}(r)$ [Eq.~(\ref{eq:pe-overline})], the familiar radiation equation of state in $d=3$. This relation applies to relativistic quarks in the nucleon. 
For baryons containing heavy quarks, however, the energy is dominated by the heavy-quark mass~\cite{Hu:2024mas}. Using the equation of motion,
$\frac{i}{4}  \bar{\psi}_H \gamma^{\{0}\!\stackrel{\leftrightarrow}{D}\!{}^{0\}}\psi_H = -i \bar{\psi}_H (\bold{D} \cdot \alpha) \psi_H + m_H \bar{\psi}_H\psi_H$, Eq.~(\ref{eq:traceless_E}) and Eq.~(\ref{eq:trace_E}) reveal that the hadron energy is $E_0 \approx m_H$. The gravitational form factors, Eqs.~(\ref{eq:Ebar}) and (\ref{eq:Etr}), then imply that 
$E(q^2)$ is dominated by $G_m(q^2)$. By contrast, the heavy-quark sigma-term contributions to the pressure cancel between Eqs.~(\ref{eq:petr}) and (\ref{eq:pe-overline}), leading to $P_H (q^2) \ll E_H (q^2)$ and correspondingly in coordinate space,
\begin{equation}  \label{eq:heavy_baryon}
    p_H(r) \ll \epsilon_H(r).
\end{equation}
%


\section{Origin of $p_{\rm tr}(r)$ and $\epsilon_{\rm tr}(r)$}

While the energy density is positive, the trace pressure-energy equation of state in Eq.~(\ref{eq:petr}) implies a negative pressure, in contrast to the positive pressure of matter and radiation in Eq.~(\ref{eq:pe-overline}). Such negative pressure corresponds to an inward stress, indicating mechanical binding of the system. 
It has been suggested that the trace anomaly and sigma terms in the hadron originate from depleting or decondensating the gluon and quark condensates of the QCD vacuum~\cite{Shifman:1978by, Shuryak:1978yk, Liu:2021gco, Liu:2023cse}. Owing to conformal symmetry breaking, QCD develops a gluon condensate associated with the trace anomaly. In addition, the quark condensate signals spontaneous chiral symmetry breaking, with the quark condensate as the order parameter.

Together, the gluon and quark condensates determine the structure of the vacuum energy-momentum tensor. In the vacuum
\begin{equation}
    \langle T^{\mu\nu}\rangle = \frac{1}{4} g^{\mu\nu} \langle T_{\mu}^{\mu}\rangle ,
\end{equation}
so that the vacuum energy density $\epsilon_{\rm vac} = \frac{1}{4}\langle T_{\mu}^{\mu}\rangle < 0$ and the corresponding pressure $p_{\rm vac} = -\,\epsilon_{\rm vac}$ are spatially uniform constants. The trace components  of the nucleon energy density, $\epsilon_{\rm tr}(r)$ and pressure $p_{\rm tr}(r)$ are matrix elements due to the trace anomaly and the quark-loop contributions to the sigma terms. In lattice QCD, these contributions are evaluated from the disconnected insertions of three-point correlation functions,
\begin{equation}   \label{eq:C_3}
   C_{3}({\rm DI}) = \langle G_2 O\rangle -  \langle O\rangle \langle G_2\rangle,
\end{equation}
where $G_2$ is the nucleon propagator and $\langle O\rangle$ is the vacuum expectation value of the operator. 
The second term removes the  contribution in which the operator insertion is uncorrelated with the nucleon propagator. Consequently, the extracted nucleon matrix element measures the change in the corresponding condensate induced by the presence of the nucleon. Since the vacuum expectation value of the trace of the EMT is negative, ($\langle T_{\mu}^{\mu}\rangle < 0$), a depletion of this negative vacuum condensate in the nucleon produces a positive vacuum-subtracted trace matrix element and hence a positive contribution to the nucleon mass. The magnitude of this contribution is determined by the relative size of 
the first and second terms in Eq.~(\ref{eq:C_3}), i.e. the extent to which the negative vacuum energy density is depleted in the nucleon.

To estimate the magnitude of this depletion, we consider an idealized model in which the vacuum condensates are completely 
dissolved in  a spherical region of uniform energy density. The corresponding mass contribution is
\mbox{$M_{\rm vac} = |\langle T_{\mu}^{\mu}\rangle| V_{\rm eff}$,}
where the effective volume $V_{\rm eff}$ is approximated by $V_{\rm eff} = \frac{4\pi}{3} \langle r^2\rangle_{\rm tr}^{3/2}$, with $\langle r^2\rangle_{\rm tr}$ the mean-square radius the nucleon trace anomaly distribution.  Since more than 90\% of the nucleon mass is generated by the trace anomaly, we use the lattice 
determination~\cite{Wang:2024lrm}, $\langle r^2\rangle_{tr}^{1/2} = 0.89 (10)$fm, together with the  phenomenological gluon 
condensate~\cite{Shifman:1978by}, $\frac{\alpha_s}{\pi} G^{\mu\nu}G_{\mu\nu} = 0.012 \, {\rm GeV}^4$, 
and the leading-order $\beta$ function to estimate $M_{\rm vac} \sim 750$ MeV. This is close to the trace-anomaly contribution to the nucleon mass, approximately 854 MeV, corresponding to 91\% of the nucleon mass. A similar estimate can be made for the quark condensate. Using $\langle \bar{q}q\rangle \sim - (270\, {\rm MeV})^3$ and a characteristic radius of approximately 0.8 fm, we obtain a condensate contribution of about 36 MeV to the \mbox{$\pi N$} sigma term. This is comparable to the disconnected-insertion result from lattice QCD, about 30 MeV~\cite{Gupta:2021ahb}. Together, these estimates support the picture that a substantial fraction of the nucleon mass originates from the depletion of the gluon and quark condensates of the QCD vacuum.

\subsection{Volume scaling and confinement}

Since the pressure is the functional derivative of the energy with respect to local volume variations [i.e., Eq.~(\ref{eq:p_mech})], the local pressure-energy equations of state in Eq.~(\ref{eq:petr}) imply global volume scaling. 
We shall explore the volume scaling based on these equations of state and their corresponding pressures.

Consider the total energy as an integral of the energy density with curvilinear coordinates
\begin{equation}
    E = \int d^3 x\, \sqrt{-g(x)}\, \epsilon (x)
\end{equation}
We introduce the metric to facilitate scale changes. Therefore, 
\begin{equation}
    V = \int d^3 x\, \sqrt{-g(x)}
\end{equation}
Defining the local volume density $v(x) \equiv \sqrt{-g(x)}$, then $\delta V(x) = \delta v(x)$.
Now consider the functional variation
with respect to $ \delta v(x)$
\begin{equation}
    \delta E = \int d^3 x \, \big[\epsilon(x)\delta v(x)+ v(x)\,\delta\epsilon(x)\big].
\end{equation}
Therefore,
\begin{equation}   \label{eq:delta_E}
    \frac{\delta E}{\delta v(x)} = \epsilon(x) + v(x)\, \frac{\delta \epsilon(x)}{\delta v(x)},
\end{equation}
where the second term accounts for the response of the energy density to the local change in volume. From the equation of state for the trace part of the energy density in Eq. (\ref{eq:petr}) and the definition of $p_{\rm mech}$ in Eq. (\ref{eq:p_mech}), one finds that
\begin{equation}
    \frac{\delta \epsilon_{\rm tr}(x)}{\delta v(x)}=0.
\end{equation}
In other words, $\epsilon_{\rm tr}$ does not respond to local volume variation, consistent with the fact that it reflects the uniform gluon and quark condensates in the vacuum. 
As a consequence, 
\begin{equation}
    \frac{\delta E_{\rm tr}}{\delta v(x)} = \epsilon_{\rm tr}(x).
\end{equation}
For a uniform dilation $v(x) \rightarrow \alpha v(x)$, $E_{\rm tr}$ scales as the volume,
\begin{equation}  \label{eq:Ealpha}
    E_{\rm tr} (\alpha) = \int d^3 x\, \alpha v(x) \epsilon_{\rm tr} = \alpha E_{\rm tr}(1)
\end{equation}
Similarly, 
\begin{equation}   \label{eq:Valpha}
    V = \int d^3 x\, v(x) \Rightarrow V(\alpha) = \alpha V(1)
\end{equation}
Combining Eqs. (\ref{eq:Ealpha}) and (\ref{eq:Valpha})
\begin{equation}
    \frac{E_{\rm tr}(\alpha)}{E_{\rm tr}(1)}
    = \frac{V(\alpha)}{V(1)} \Rightarrow E_{\rm tr} \propto V.
\end{equation}
Thus, $E_{\rm tr}$ scales as the volume. The mechanical pressure conjugate to the virtual volume variation is then
\begin{equation}
    P_{\rm eff, tr} = - \frac{d E_{\rm tr}(V)}{d V}
    = - \, {\rm constant}. 
\end{equation}
This constant attractive pressure as the hadron expands implies volume (radial) confinement and is the 3D extension of the linear confinement, which are familiar in heavy quarkonium cases. This is the major conclusion we obtain from the trace part of
the pressure-energy equation of state.
On the other hand, for the traceless part
\begin{equation}
 \frac{\delta \overline{E}}{\delta v(x)} = - \frac{1}{3}\, \overline{\epsilon}(x)
\end{equation}
Eq. (\ref{eq:delta_E}) becomes
\begin{equation} \label{eq:traceless_v(x)}
    v(x) \, \frac{\delta \overline{\epsilon}(x)}{\delta v(x)} = - \frac{4}{3} \,\overline{\epsilon}(x).
\end{equation}
For a uniform dilation $v(x) \rightarrow \alpha v(x)$ everywhere, then Eq. (\ref{eq:traceless_v(x)}) becomes an ordinary differential equation. Upon integration
\begin{equation}   \label{eq:v4/3}
    \frac{d \overline{\epsilon}}{\overline{\epsilon}} = - \frac{4}{3} \frac{d\,v}{v} \Rightarrow \overline{\epsilon} \propto v^{-4/3}.
\end{equation}
Since $V = \int d^3 x\, v \propto v$, Eq. (\ref{eq:v4/3}) is
\begin{equation}
    \overline{\epsilon} \propto V^{-4/3}
\end{equation}
From $\overline{E} = \int d^3 x \,v \,\overline{\epsilon}$, the volume scaling of $\overline{E}$ is
\begin{equation}
\overline{E} \propto V^{-1/3}
\end{equation}
The corresponding pressure is then
\begin{equation}    \label{eq:P_tl}
    \overline{P}_{\rm eff} = - \frac{d \overline{E}(V)}{d V}
    = c\, V^{-4/3}
\end{equation}
which is positive and decreases with volume. 

   The volume scalings of $ E_{ \rm tr}$ and $\overline{E}$ are analogous to the MIT bag model, where the bag constant gives
   a linear V dependence of the rest energy and the quark and gluon energies are inversely proportional to the bag radius R (or
   $V^{-1/3}$). The equilibrium volume $V_0$ is determined by the pressure balance condition $d E_0/dV = 0$, which is just the
   virial theorem~\cite{Liu:2023cse}. By analogy, we can determine $P_{\rm eff, tr}$ by defining $P_{\rm eff, tr} V_0 =  - \int  d^3 r \,\epsilon_{\rm tr} (r) = - M/4$ with the equilibrium volume $V_0$  estimated from the spatial distribution $\widetilde{G}_m(r)$. A similar definition can be obtained for $\overline{P}_{\rm eff} $ in Eq.~(\ref{eq:P_tl}). At the equilibrium volume $V_0$, 
$\overline{P}_{\rm eff}  = - P_{\rm eff, tr}$.

Since $p = - \epsilon$ is a signature property for the condensate, this suggests that the presence of condensates generates a confining mechanism--to the extend that the vacuum condensate is largely depleted by the hadron, the total pressure is balanced between the dynamical pressure from quarks and gluons and the static, vacuum-induced pressure from the depletion of the condensates. This is consistent with the conclusion based on volume scalings from the total pressure-volume work and energy relations~\cite{Liu:2021gco,Liu:2023cse}. The force and its relation to normal and shear stresses are given in Appendix A.

We should note that this volume confinement is distinct from color confinement mechanisms based on monopoles or center vortices~\cite{Greensite:2003bk, Shifman:2010jp, Brodsky:2014yha}. Rather, it represents a three-dimensional manifestation of linear confinement, as realized, for example, in charmonium. Within a renormalization group framework, the Wilson-loop potential and its derivative can be related to the trace anomaly~\cite{Dosch:1995fz, Rothe:1995hu}
\begin{equation}  \label{plaquette}
V(r) + r \frac{\partial V(r)}{\partial r}  =  \lim_{T \rightarrow \infty} \frac{\langle \frac{\beta}{2g} \int d^3\vec{x}\, (- E^2)\, W_L(r,T)}{\langle W_L(r,T)\rangle} + \lim_{L \rightarrow \infty} \frac{\langle \frac{\beta}{2g} \int d^3\vec{x}\,  B^2\, W_L(r,L) \rangle}{\langle W_L(r,L)\rangle},
\end{equation}
where $W_L(r,T)/W_L(r,L)$ denote the space-time and space-space Wilson loops with spatial separation $r$. 
Lattice simulations reveal a linear potential between infinitely heavy quark-antiquark pairs from Eq.~(\ref{plaquette})~\cite{Dosch:1995fz}, consistent with the Wilson-loop area law~\cite{Bali:1997am}, and the formation of a flux tube between separated color sources~\cite{Baker:2018mhw}. These results support a picture of a constant vacuum energy density  $\langle T^{00}\rangle$. It is found in a lattice simulation of chromo field distribution between two static color sources~\cite{Bali:1994de}  that after a distance of $\sim$ 0.75 fm, a flux tube of the action density is formed with a constant 
cross sectional area $A$, the static potential between the heavy quark-antiquark pair then takes the form
\begin{equation}
V(r)=\epsilon_{\rm vac}\, A \, r = \sigma r,
\label{linear_potential}
\end{equation}
which gives rise to the linearly increasing potential, with $\sigma$ the string tension~\cite{Bali:1994de}.

\section{Vortex of type-II superconductor} 

It was pointed out~\cite{Liu:2023cse} that hadrons share a close analogy with vortices in type-II superconductors in terms of their energetics and pressure-energy relations. In a vortex, the external magnetic field penetrates the core and decays radially with the London penetration depth $\lambda_L$. The local density of superconducting electrons, $n_c$, vanishes in the normal-phase core and recovers to its bulk value over the coherence length $\xi$. The system is type II when the Ginzburg-Landau parameter satisfies $\kappa = \frac{\lambda_L}{\xi} > \frac{1}{\sqrt{2}}$.

Type-II superconductors are described by the Ginzburg-Landau equation, which determines the superconducting order parameter $\psi(r)$, with the local condensate density given by $n_c = |\psi(r)|^2$. We focus here on the energetics, in particular the origin of the various contributions to the vortex-core energy relative to the Meissner state~\cite{Clem:1975}.
The Ginzburg-Landau energy of a vortex contains several contributions that admit a one-to-one correspondence with those in a hadron. The magnetic field energy is analogous to the chromoelectric and chromomagnetic field energies in Eq.~(\ref{eq:traceless_E}). The supercurrent energy corresponds to the quark kinetic and potential energies in Eq.~(\ref{eq:traceless_E}). Finally, the energy cost of depleting the Cooper-pair condensate, characterized by the lower critical field $H_{c1}$, parallels the trace-anomaly and sigma-term contributions in Eq.~(\ref{eq:trace_E}), representing the energy required to deplete the vacuum gluon and quark condensates inside the hadron.

A variational approach is employed~\cite{Clem:1975}, in which the order parameter is assumed to take the form, $\Psi(\rho, \phi) = f(\rho) e^{-i\phi}$ 
 and $f(\rho) = \rho/\sqrt{\rho^2 + R^2}$, where $\rho$ is the cylindrical radial coordinate and $R$ is a variational parameter characterizing the core radius. The condensate density is then given by $n_c = |\Psi|^2 = f(\rho)^2$. The free energy per unit length of the vortex line, expressed in units of $\phi_0 H_c/(2\sqrt{2}\pi)$, is obtained as 
\begin{equation}   \label{eq:free_energy}
F' \equiv \frac{F}{l\,\phi_0 H_c/(2\sqrt{2}\pi)}= \frac{1}{8} \kappa R'^2 + \frac{1}{8\kappa} + \frac{K_0(R')}{2\kappa\,K_1(R') R'},
\end{equation}
where $R' = R/\lambda_L$ and $K_0$ and $K_1$ are modified Bessel functions of the second kind. The first term represents the cost of depleting the condensation and is proportional to the core area. The second and third terms arise from the supercurrent and magnetic field energies.
The equilibrium condition, obtained by minimizing the energy with respect to the area, $\frac{dF'}{dA}A = 0$, determines R to be of order $\lambda_L$. For $R' \sim 1$, the ratio $K_0(R')/K_1(R')$ varies slowly, and the third term is effectively dominated by the 
$1/R'$ dependence. In this regime, condensate depletion produces a constant two-dimensional free-energy density and hence a constant negative pressure.  This is balanced by the positive dynamical pressures arising from the magnetic field and supercurrent contributions, which satisfy a radiationlike equation of state $p=\epsilon/d$ with $d=2$. Therefore,  
\begin{equation}   \label{eq:vortex}
    p_{\rm sta} = - \,\epsilon_{\rm sta},  \hspace{1cm} 
    p_{\rm dyn} (R') = \frac{1}{d}\, \epsilon_{\rm dyn} (R') + ...
\end{equation}
The ellipsis represents small contributions from the terms that deviate from $1/R'$ in Eq. (\ref{eq:free_energy}).


The underlying physics of vortices and hadrons is fundamentally different. In superconductivity, the Cooper-pair condensate spontaneously breaks a global 
$U(1)$ symmetry and generates a gauge-boson mass through the Anderson-Higgs mechanism. In contrast, the gluon condensate in QCD arises from the explicit breaking of conformal symmetry via the trace anomaly and is not associated with spontaneous symmetry breaking or gluon mass generation, while the quark condensate is the order parameter of the spontaneous chiral symmetry breaking. Nevertheless, their static pressure-energy equations of state are identical because of the presence of condensates. In this sense, the volume-confinement mechanism operates analogously in the two systems.

\section{Cosmological constant}

It has been pointed out that the trace matrix element bears a close analogy to the cosmological constant~\cite{Liu:2021gco, Liu:2023cse}. Both are associated with the metric $g^{\mu\nu}$ in the context of the energy-momentum tensor and, therefore, obey the same pressure-energy relation. In particular, the pressure equals the negative of the corresponding energy density as a consequence of the metric structure.

To solve the Friedmann equations, one needs to specify the equations of state relating pressure and energy density for the different components of the Universe. In the standard $\Lambda {\rm CDM}$ cosmological model, these equations of state are given by 
 \begin{equation}
     p = \omega\, \epsilon
 \end{equation}
where $\omega$ is the proportionality constant. As discussed above, $\omega = -1$ for the cosmological constant.\footnote{Recent baryon acoustic oscillation  data from the Dark Energy Spectroscopic Instrument have raised tensions regarding the constancy of the cosmological constant--this issue is beyond the scope of the present work.} This is identical to the pressure-energy equation of state  for the trace component in QCD and  the condensate-depletion contribution in a superconducting vortex.

For radiation and relativistic neutrinos, $\omega = 1/d$ (with $d = 3$ in this case). This is identical to the relation for relativistic quarks and gluons in hadrons [Eq.~(\ref{eq:pe-overline})], as well as for the electrons and magnetic field in a superconducting vortex [Eq.~(\ref{eq:vortex})]. For nonrelativistic matter, $\omega \approx 0$, which likewise parallels the case of heavy-quark hadrons [Eq.~(\ref{eq:heavy_baryon})].

The physics of gravity and gauge theory is fundamentally different. In hadrons and vortices, negative pressure corresponds to an inward stress and hence attraction, whereas in cosmology it drives accelerated expansion, effectively acting as repulsion. Nevertheless, their pressure-energy equations of state are formally identical. This raises a natural question: since the $p = - \,\epsilon$ part of pressure-energy relation in hadrons and vortices originates from condensates, could the cosmological constant likewise arise from a condensate? We leave this possibility for future investigation.

\section{Summary}

By identifying the mass form factor obtained from the matrix elements of the trace of the EMT with that defined through the gravitational form factors, one can express the pressure form factor $D(q^2)$ in terms of $A(q^2), B(q^2)$, and the mass form factor $G_m(q^2)$, once EMT conservation is taken into account. This, in turn, leads to the pressure-energy form factor relations for both the traceless and trace parts of the EMT. 

Interpreting the local mechanical pressure--namely, the matrix element of $\frac{1}{3} T^{ii}(r)$--as the stress defined by the functional derivative of the nucleon energy with respect to the local volume, we demonstrate that the trace part of the energy scales linearly with the volume. This supports the interpretation that the trace part of the energy acts as a potential arising from the depletion of the gluon and quark condensates inside the nucleon. The associated static pressure is a negative constant, leading to a mechanism of volume confinement, analogous to the linear potential in charmonium. In contrast, the traceless part of the dynamical energy associated with quark and gluon fields scales as $V^{-1/3}$, leading to a positive pressure in Eq.~(\ref{eq:pe-overline}).
There are several ways to
decompose the nucleon rest energy~\cite{Ji:1994av,Lorce:2017xzd, Metz:2020vxd,Liu:2023cse}, it is the decomposition of $T^{00}$ and $T^{ii}$ 
into their traceless and trace components that establishes the pressure-energy equations of state, thereby clarifying their origins and physical significance.

It is noteworthy that vortices in type-II superconductors obey the same equations of state as the nucleon. 
The common feature shared by hadrons and vortices is that their stability originates from the presence of condensates, which generate a negative constant pressure that provides confinement.

In addition, we point out that the standard cosmological model, $\Lambda {\rm CDM}$, with a cosmological constant, obeys the same set of equations of state. The equation of state for nonrelativistic matter is likewise analogous to that of a hadron containing heavy quarks. Given that the cosmological constant plays a role similar to the trace anomaly in QCD, it has been speculated that, as in QCD, it may originate from a condensate~\cite{Liu:2021gco,Liu:2023cse}. This intriguing possibility will be explored in future work.

\section{ACKNOWLEDGMENTS}
The author is indebted to  S. Brodsky,  M. Chanowitz, X. Ji, V. Koch, C.S. Lam, G. Miller, J.C. Peng, \mbox{M. Peshkin}, F. Yuan, and R. Zwicky  for fruitful discussions. He also thanks D. Pefkou for providing the
lattice data and \mbox{F. He} for help with the figures. This work is partially supported by the U.S.
DOE Grant No. DE-SC0013065 and U.S. DOE Office of Nuclear Physics under the umbrella of Quark-Gluon Tomography Topical Collaboration with Award No. DE-SC0023646. 

\section{DATA AVAILABILITY}

There are no publicly available research data or software supporting this manuscript. Requests for further information or
data should be sent to the author.

\bibliographystyle{unsrt}
\bibliography{References}

\begin{center}
APPENDIX A: SHEAR STRESS AND FORCE
\end{center}

 We shall discuss the force in relation to the normal stress (pressure) and shear stress. The spatial distribution from the $ \langle\langle T^{ij}\rangle\rangle$ form factor is~\cite{Polyakov:2018zvc}
\begin{equation}
    \langle\langle T^{ij}\rangle\rangle(r)
   = (\frac{r^ir^j}{r^2} - \frac{1}{3} \delta^{ij})s(r) + \delta^{ij} p(r),
\end{equation}
where, $p(r)$ is the normal stress and $s(r)$ is the shear stress. Defined from the Fourier transform $\widetilde{D}(r)$, they are
\begin{eqnarray}
 p(r) & =& \frac{2}{3M}\frac{1}{r^2} \frac{d}{dr} r^2 \frac{d}{dr} \widetilde{D}(r) \nonumber \\
 s(r) &=& - \frac{1}{M} r \frac{d}{dr} \frac{1}{r} \frac{d}{dr} \widetilde{D}(r).
 \end{eqnarray}
Replacing $q^2 D(q^2)$ in terms of the $A(q^2), B(q^2)$  and $G_m(q^2)$ form factors in 
Eq.~(\ref{eq:D(q^2)}), 
\begin{eqnarray}
p(r) & =& \frac{2M}{9}\int \frac{d^3q}{(2\pi)^3} e^{-i \vec{q}\cdot\vec{r}} P_0({\rm cos\theta})\Big [A(q^2) - G_m(q^2)  + \frac{q^2B(q^2)}{4M^2} \Big ] \nonumber \\
s(r) & =& \frac{M}{2}\int \frac{d^3q}{(2\pi)^3} e^{-i \vec{q}\cdot\vec{r}} P_2({\rm cos\theta})\Big [A(q^2) - G_m(q^2)  + \frac{q^2B(q^2)}{4M^2} \Big ].
\end{eqnarray}
This gives the expression of $p(r)$ in Eqs.~(\ref{eq:pr_overline}) and (\ref{eq:prtr}) in terms of
$\widetilde{A}(r), \widetilde{B}(r)$ and $\widetilde{G}_m(r)$.

Due to the conservation of EMT in the static case,
$\bigtriangledown^{i}T_{ij} = 0$, one has the differential equation~\cite{Polyakov:2018zvc}
\begin{equation}   \label{eq:p-s relation}
    p'(r) + \frac{2}{3}s'(r) + \frac{2}{r} s(r) = 0.
\end{equation}
In view of spherical symmetry, the pressures in the local spherical coordinates $(\hat{r}, \hat{\theta}, \hat{\phi})$ are diagonal 
\begin{equation}
    T^{ab} = {\rm diag}(p_r, p_t, p_t),
\end{equation}
where the radial pressure $p_r(r)$  and tangential pressure $p_t(r)$ are~\cite{Lorce:2018egm}
\begin{equation}
    p(r) = \frac{p_r(r) + 2 p_t(r)}{3}, \hspace{1cm} 
    s(r) = p_r(r) - p_t(r),
\end{equation}
Therefore, the shear stress $s(r)$ is actually the pressure anisotropy, instead of the truly
off-diagonal shear.
In this case, Eq.~(\ref{eq:p-s relation}) becomes
\begin{equation}
    p'_r(r) + \frac{2s(r)}{r} = 0.
\end{equation}
Several integral relations can be derived~\cite{Polyakov:2018zvc,Lorce:2018egm}. Among them is the von Laue condition in Eq.~(\ref{eq:vonLaue}). 

One can define the differential force $dF^i$ from the spatial part of the EMT with the $T^{ij}$ acting on the infinitesimal surface $dA$ normal to the $j$ direction, i.e.
\begin{equation}
    dF^i(r)=  T^{ij}(r) n^j dA,
\end{equation}
where $n^j = r^j/r$ and we have simplified  $\langle\langle T^{ij}\rangle\rangle(r)$ as 
$T^{ij}(r)$. Since $T^{ij}(r)$ has the form
\begin{equation}
    T^{ij}(r) = \delta^{ij} p(r) + (\frac{r^ir^j}{r^2} - \frac{1}{3} \delta^{ij})s(r),    
\end{equation}
\begin{equation}   \label{eq:force_p_s}
    \frac{dF^i(r)}{dA} = (p(r) + \frac{2}{3} s(r)) n^i = p_r(r)n^i.
\end{equation}
It has been argued~\cite{Perevalova:2016dln} that the force should be directed outward in order to avoid collapse of the system.  Indeed, it is found that
$p(r) + \frac{2}{3}s(r) > 0$ in a recent lattice calculation~\cite{Hackett:2023rif}. This shows that the force does not change sign, in contrast to the pressure $p(r)$, which means that for the region of $r$ that $p(r)$ is negative due to the static pressure from the trace part in Fig.~\ref{fig:pr} is filled up by the $s(r)$.

While the force is directed outward, its integration over any closed surface vanishes. The surface integral
\begin{equation}
    F^i = \oint_S T^{ij} n^j dA = \int_V \partial_{j}T^{ij} d^3\,r = 0
\end{equation}
becomes a volume integral due to the Gauss law and it vanishes because of conservation of EMT for a static, spherical bound system. There is no net force on any surface. This also proves why $p_r(r)$ does not change sign. Finally, one can consider the traceless and trace parts of the force through 
$p(r)$ and $s(r)$ in Eq.~(\ref{eq:force_p_s}).

\end{document}